\documentclass[%
 reprint,
 amsmath,amssymb,
 aps,
]{revtex4-2}

\usepackage{graphicx}
\usepackage{dcolumn}
\usepackage{bm}

\usepackage{multirow}
\usepackage{hyperref}

\begin{document}

\preprint{APS/123-QED}

\title{Improving target neutron momentum reconstruction using MINER$\nu$A $\pi^0$ data}%

\author{R K Pradhan\textsuperscript{1}}%
 \email{kumarriteshpradhan@gmail.com}

\author{R Lalnuntluanga\textsuperscript{1,2}}
\email{tluangaralte.phy@gmail.com}

\author{A Giri\textsuperscript{1}}%
 \email{giria@phy.iith.ac.in}
 
\affiliation{%
\textsuperscript{1}
 \textit{Indian Institute of Technology Hyderabad, Hyderabad, 502284, Telangana, India}
}%

\affiliation{\textsuperscript{2}
\textit{Tel Aviv University, Tel Aviv 69978, Israel}
}


\begin{abstract}
With the neutrino experiments advancing toward high-precision measurements and greater emphasis on reducing systematic uncertainties, improving the single-pion production models, a major component of the hadronic activity observed in the neutrino oscillation experiments, into the Monte Carlo simulations is crucial. This work presents the predictions of the struck nucleon’s Fermi motion by analyzing the charged-current neutral pion production on carbon nucleus in MINER$\nu$A. A minimal variation of GENIE and NuWro based on their default models shows an improvement in the prediction of single $\pi^0$ production. The prediction describes the data more accurately in the higher-momentum tail; however, discrepancies between the predictions and data below the Fermi peak highlight the limitations in current nuclear models used in the Monte Carlo generators.
\end{abstract}

\maketitle


\section{\label{sec1}Introduction}
The precision physics of neutrino experiments at a few GeVs is hindered by the neutrino-nucleus interactions information \cite{NuSTEC:2017hzk}. The requirement for high-statistics data leads to a challenge due to the complex effects of the nuclear medium, which contribute significantly to systematic uncertainties in neutrino cross section measurements and the determination of the charge-parity violation parameter \cite{NOvA:2019cyt, T2K:2019bcf}. Neutrinos interact with the nucleons inside the nuclear medium, and the produced hadrons can reinteract with the nucleons within the nucleus, which is called final-state interactions(FSIs). Because of FSIs, pions can be produced, causing quasielastic (QE) interactions to be misidentified as non-QE, or pions produced in the resonance (RES) channel can be absorbed inside the nucleus, leading to their misidentification as QE. Neutrinos can interact with the nucleon pairs bound by pion exchange, leading to a multinucleon knockout known as the meson exchange (MEC) process. Neutrino experiments rely extensively on the Monte Carlo models to predict the neutrino-nucleus interactions for physics studies. The production of pions in the neutrino interaction is one of the critical channels for the DUNE \cite{DUNE:2015lol,Lalnuntluanga:2023qkp,Lalnuntluanga:2024ssd} experiment, and important as well in the atmospheric neutrino experiments such as JUNO \cite{JUNO:2015zny}, Hyper-K \cite{Hyper-KamiokandeProto-:2015xww} and Super-K \cite{Super-Kamiokande:1998kpq}. Therefore, understanding the nuclear effects in the pion production cross section of the neutrino-nucleus interaction is vital for high-precision measurement of physics parameters in the neutrino experiments.

In this work, we study the uncertainties associated with the single neutral pion production in the charge current neutrino-carbon interactions using  Monte Carlo models: GENIE \cite{andreopoulos2015genieneutrinomontecarlo} and NuWro \cite{Benhar:2005dj}. We aim to validate the initial neutron momentum, which is known as the Fermi motion of nucleons in the nuclear potential. The transverse component of the momentum can be reconstructed using the transverse kinematic imbalance(TKI) concept \cite{Lu:2015tcr}, which provides a precise approach to identifying nuclear effects or their absence \cite{MINERvA:2018hba,MINERvA:2019ope,Furmanski:2016wqo,T2K:2018rnz,Cai:2019jzk}. The Monte Carlo predictions are obtained using the recent MINER$\nu$A $\pi^0$ measurement \cite{MINERvA:2020anu} by varying models and their parameters. Previous studies \cite{MINERvA:2020anu} using NuWro and GiBUU \cite{Buss:2011mx} Monte Carlo models show the discrepancy in the generator predictions of the $p-\pi^0$ cross section on neutron inside nuclei using the Fermi Gas as well as the spectral function despite its success in describing the QE-like results \cite{MINERvA:2018hba,T2K:2018rnz}. Recent works on this problem have been done in NuWro \cite{Yan:2024kkg} and GENIE \cite{GENIE:2024ufm} focusing on tuning the models and various parameters, as well as incorporating hybrid models to improve the predictions. The motivation for this work is to explore the existing Monte Carlo models, and identify potential physics-based modifications that can improve their accuracy in describing the data. The methods for investigating the nuclear effects are discussed in Sec. \ref{sec2}, and the details of the models used in the simulation and analysis are explained in Sec. \ref{sec3}. The results and conclusions from our analysis are presented in Secs. \ref{sec4} and \ref{sec5} respectively.

\section{\label{sec2}Target Nucleon Momentum Reconstruction}

The momentum of the struck nucleon can be reconstructed using the kinematics of the final-state particles. The transverse component is a TKI variable that shows an imbalance in momentum conservation due to nuclear effects \cite{Lu:2015tcr}. The longitudinal component of the momentum can be reconstructed using energy and momentum conservation, as described below \cite{Furmanski:2016wqo}.



Consider a charged current (CC) $\nu_{\mu}$ interaction with the Carbon nucleus as follows:
\begin{equation*}
    \nu_{\mu} \,+\, C \rightarrow \mu^{-}\,+ p\,+ \pi^0 \,+ X
\end{equation*}
X is the hadronic system consisting of nuclear remnant and possible additional protons or $\pi^0$'s without any other mesons. The transverse momentum ($\delta p_T$) and the initial-state nucleon momentum ($p_N$) can be calculated as follows \cite{MINERvA:2018hba}. 
\begin{equation}
    \vec{\delta p_T} = \vec{p}_T^{\,\mu} + \vec{p}_T^{\,h}
\end{equation}
Here, $\vec{p}_T^{\,h} = \vec{p}_T^{\,p}+ \vec{p}_T^{\,\pi^0}$. T refers to the transverse component of the respective momentum. From the conservation of energy and the longitudinal component of momentum, 
\begin{equation*}
    E_{\nu_{\mu}} + m_C = E_{\mu} + E_h + E_{C'} 
\end{equation*}
\begin{equation*}
    E_{\nu_{\mu}} = p_L^{\mu} + p_L^{h} - \delta p_L
\end{equation*}
The above two equations can be solved for $\delta p_L$,
\begin{equation*}
    \delta p_L = \frac{R}{2} - \frac{m_{C'}^2 + \delta p_T^2}{2R}
\end{equation*}
$R \equiv m_C + p_L^{\mu} + p_L^h - E_{\mu} - E_h$, where L represents the longitudinal momentum component to neutrino direction. $E_{\mu}$ and $E_h$ are the energies of muon and hadrons, respectively. $m_{C'}$, the mass of the carbon nucleus after the interaction, is given by $m_{C'} = m_C - m_n +b$, where $m_C$ is the carbon nucleus mass before interaction, $m_n$ is the neutron mass, and b is the excitation energy, which is taken as 28.7 MeV \cite{Furmanski:2016wqo} for carbon. Now, the momentum of the initial nucleon is defined as
\begin{equation}\label{pn}
    p_N = \sqrt{\delta p_T^2 + \delta p_L^2}
\end{equation}
$p_N$ can be beyond the Fermi level because of the extra final-state particles X due to FSIs and other intranuclear dynamics. $p_N$ is sensitive to both initial- and final-state nuclear effects.

\section{\label{sec3} Simulation Details}

\subsection{Monte Carlo event generators}
\begin{figure}[!h]
    \centering
    \includegraphics[width=9.5cm,height=7cm]{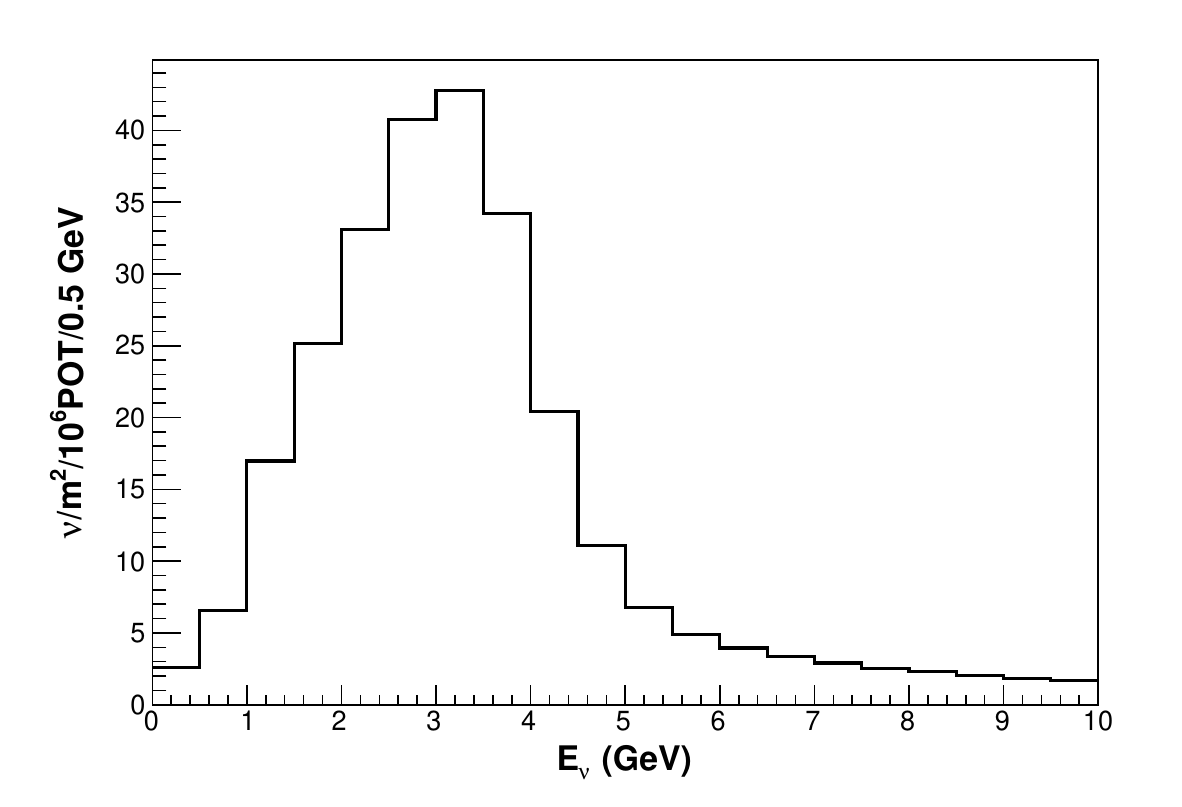}
    \caption{Low-energy (LE) muon neutrino flux for MINER$\nu$A detector in neutrino mode.}
    \label{fig:1}
\end{figure}

Neutrino event generators are indispensable tools for the precise modeling and quantitative analysis of neutrino scattering processes, serving as the foundation for a wide range of research applications in this field. This work uses GENIE \cite{andreopoulos2015genieneutrinomontecarlo} and NuWro \cite{Juszczak:2005zs} to simulate neutrino-nucleus interactions and cross section calculations. One million neutrino events with the carbon nucleus are simulated using the MINER$\nu$A low-energy flux that peaks around 3 GeV \cite{MINERvA:2016iqn} as shown in Fig.\ref{fig:1} considering the charged current (CC) Quasi-elastic (QE), Resonance (RES), Deep inelastic (DIS), Meson exchange (MEC), and Coherent (COH) channels.

\subsubsection{GENIE}

GENIE v3.01.00 with the tune G18\_10a\_02\_11a is used in this work. The local Fermi-gas model (LFG) and the spectral function (SF) \cite{Benhar:1994hw} are used to describe the nuclear structure. GENIE considers the nucleon-nucleon correlation \cite{Cruz-Torres:2017sjy} with a momentum cutoff of 0.7 GeV for LFG and the removal energy of 0.025 GeV for the carbon nucleus. The models used for QE, MEC, RES, and DIS are the Valencia QE model \cite{Gran:2013kda}, the Valencia MEC model \cite{Nieves:2016sma}, the Berger-Sehgal (BS) model \cite{Berger:2007rq}, and the Bodek-Yang model \cite{Bodek:2002vp}, respectively. All 17 resonances are considered in this simulation with an axial mass of 1.065 GeV. The invariant hadronic mass threshold for the RES-DIS joining scheme is taken as 1.8 GeV. The hN model \cite{GENIE:2021npt} is used for the FSI effects in this work. RES and DIS dominate pion production, making the RES axial mass a key parameter for the analysis. This work focuses on the momentum of target nucleons, which is described by nuclear models. A fraction of nucleons are in correlated pairs due to the short-range component of the nuclear force, referred to as the short-range correlation (SRC) fraction. SRC affects the nuclear initial state by introducing a high-momentum tail in the nucleon momentum distribution \cite{Meng:2023age}. By default, GENIE assigns 20\% of nucleons to be part of SRC pairs in heavy nuclei \cite{Guo:2021zcs, CLAS:2005ola}. This analysis focuses on the impact of the SRC fraction parameter.


\subsubsection{NuWro}

The latest version of NuWro, v21.09.2, is used to calculate cross-sections and simulate neutrino events. The initial nuclear state is described using the LFG and SF models. The Llewellyn-Smith (LS) model \cite{LlewellynSmith:1971uhs}, the Rein-Sehgal (RS) model \cite{Rein:1980wg}, and the BS model are used to simulate QE, $\Delta$(1232) resonances, and DIS, respectively. The RES-DIS transition threshold is set at 1.8 GeV. A cascade model, based on the algorithm by Metropolis \textit{et al.} \cite{Metropolis:1958sb}, is used to describe FSI. Formation zone (FZ) effects \cite{Golan:2012wx} are included in NuWro hadrons produced inside the nuclear medium. The produced hadron does not immediately interact with the nuclear medium due to the time needed for its formation. The FZ effect is relevant in DIS, RES, and FSI processes. This effect is important for pion production, as it delays the time for the pion to fully form, reducing the probability of early absorption and allowing more pions to escape the nucleus. The FZ effect models considered for FSI are the coherence length \cite{Battistoni:2009zzb} for QE and the model based on $\Delta$ lifetime for resonance pion production \cite{Golan:2012wx}. For DIS, the model is based on hadron-hadron and hadron-nucleus collision \cite{Ranft:1988kc}, and the FZ effect is off for MEC.

\subsection{MINER$\nu$A selections}

The selected CC $\nu_{\mu}$ events on the carbon nucleus are defined as those whose final state consists of $1\mu^-1p1 \pi ^0$, with any number of neutrons. The kinematic acceptance for particle detection (detector threshold) is as follows \cite{MINERvA:2020anu}:
\begin{itemize}
    \item muon: $1.5 \leq p_{\mu} (GeV/c) \leq 20.0$, $\theta_{\mu} < 25^ \circ$
    \item  proton: $p_p \geq 0.45$ GeV/c
\end{itemize}
where p and $\theta$ are the momenta and polar angle to neutrino direction, respectively. The MC predictions are compared with the deconvoluted cross section data obtained from Fig. 5 in Ref. \cite{MINERvA:2020anu}.

\section{\label{sec4} Results and Analysis}

\subsection{Comparison data: MC using untuned generators}

\begin{figure*}
    \centering
    \includegraphics[width=19cm,height=7cm]{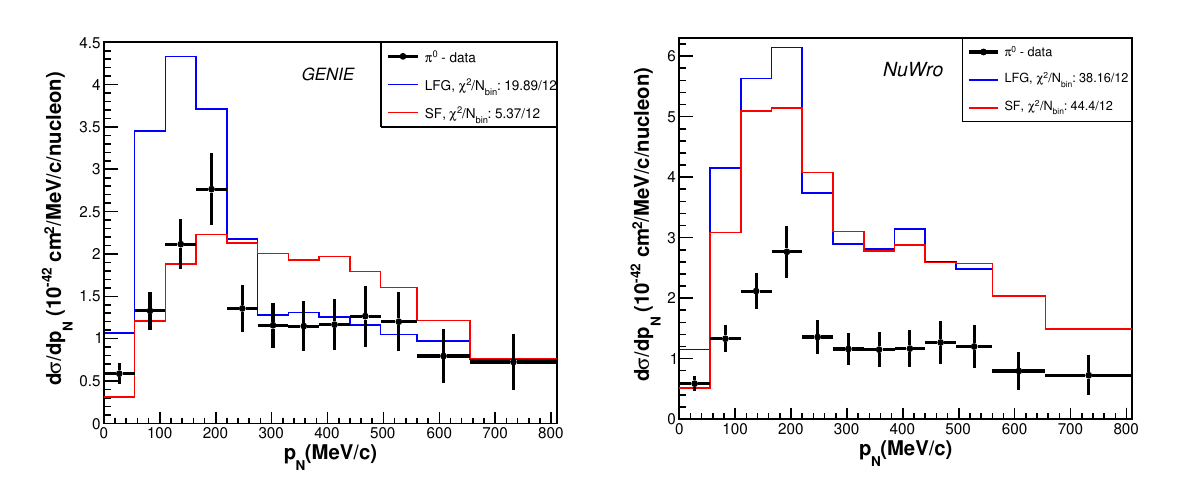}
    \caption{The flux-integrated cross section in $p_N$ from MINER$\nu$A $\pi^0$ data compared to the prediction for single $\pi^0$ production by GENIE (left) and NuWro (right) using default parameters. }
    \label{fig:3}
\end{figure*}

\begin{figure*}
    \centering
    \includegraphics[width=19cm,height=6.5cm]{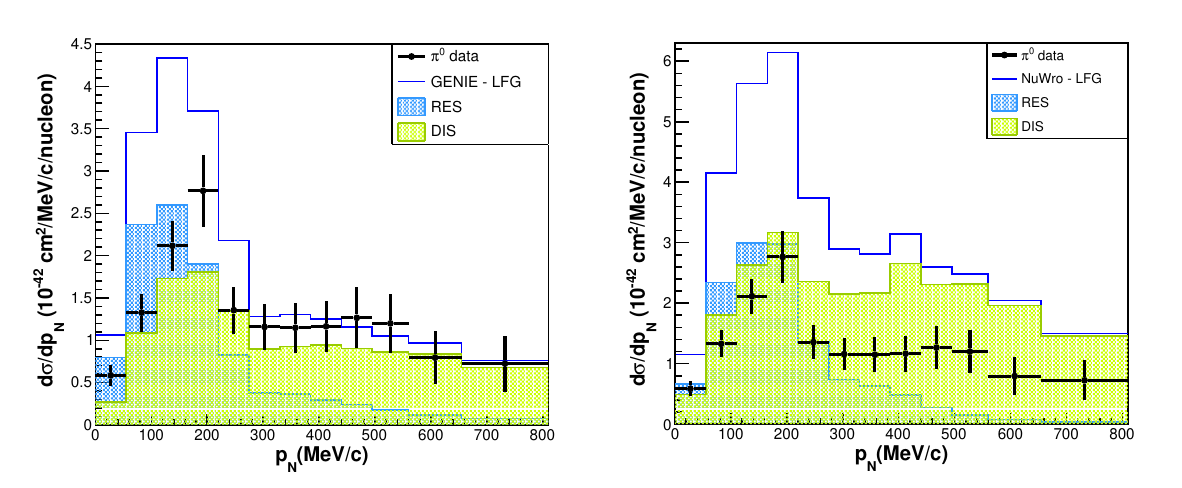}
    \includegraphics[width=19cm,height=6.5cm]{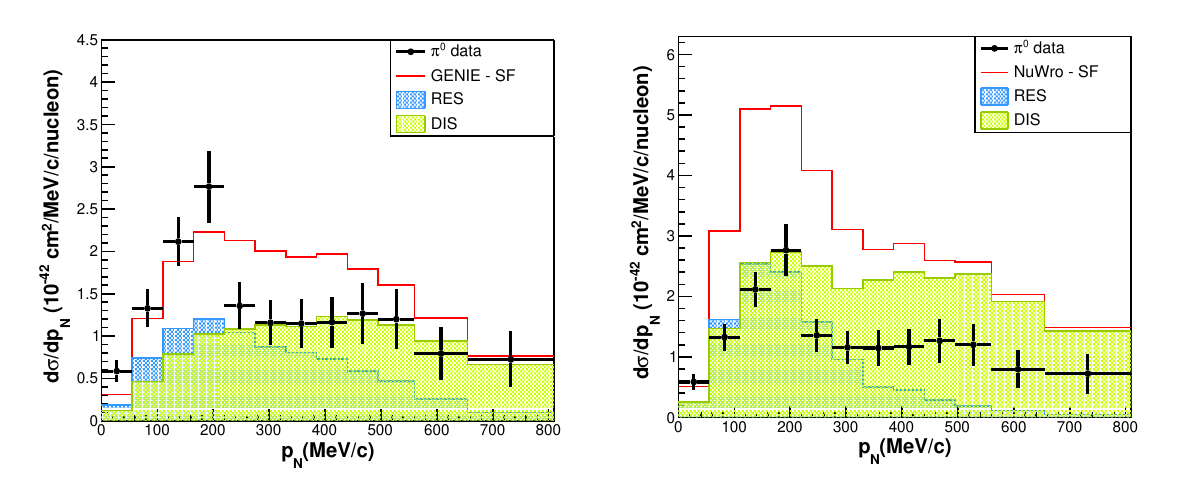}
    \caption{Contributions of RES and DIS for default LFG (top) and SF (bottom) to the GENIE (left) and NuWro (right) predictions.}
    \label{fig:new}
\end{figure*}

The momentum of the initial neutron was reconstructed using Eq. (\ref{pn}) by selecting the events with $1\mu^-1p1 \pi ^0 Xn$ in the final state. The cross section in $p_N$ for nuclear model LFG and SF using GENIE is shown in the left panel of Fig.\ref{fig:3}. For the SF, the Fermi peak is the same as the data ($\approx 200$ MeV) whereas the Fermi peak for LFG is lower.  The SF agrees with the data for the lower values of $p_N$ below the Fermi peak but for $p_N$ above the Fermi peak, the GENIE prediction is higher than the data. The LFG has a higher prediction for lower values of $p_N$ but has a good agreement in the higher tail. The dip in cross section at $p_N\,\approx$ 350 MeV/c is missing for both LFG and SF. The $\chi^2$ per bin for SF (5.37/12) is lower than that for LFG (19.89/12), but LFG provides a better shape prediction than SF. The contributions from RES and DIS to the predictions are shown in Fig.\ref{fig:new}. For SF, there is no room for RES at higher values of $p_N$, as the DIS prediction closely matches the data in that region, resulting in an overprediction above the Fermi peak. In contrast, for LFG, RES aligns well with the data at lower values of $p_N$, leaving no room for DIS, which leads to an overprediction below the Fermi peak. 

The right panel of Fig.\ref{fig:3} shows the prediction for NuWro with FSI on for SF and LFG. Both LFG and SF have the Fermi peak as the same as the data ($\approx 200$ MeV). The prediction is higher than the data by a factor of $\approx$ 0.5; however, the shape of the prediction matches the data for LFG. Also, the $\chi^2$ per bin for LFG (38.16/12) is lower than that for SF (44.4/12) therefore, LFG is considered for further analysis. Figure \ref{fig:new} shows the extra factor in NuWro prediction is due to the greater contribution of DIS events to the neutral pion production.

\begin{figure*}
    \centering
    \includegraphics[width=19cm,height=7.5cm]{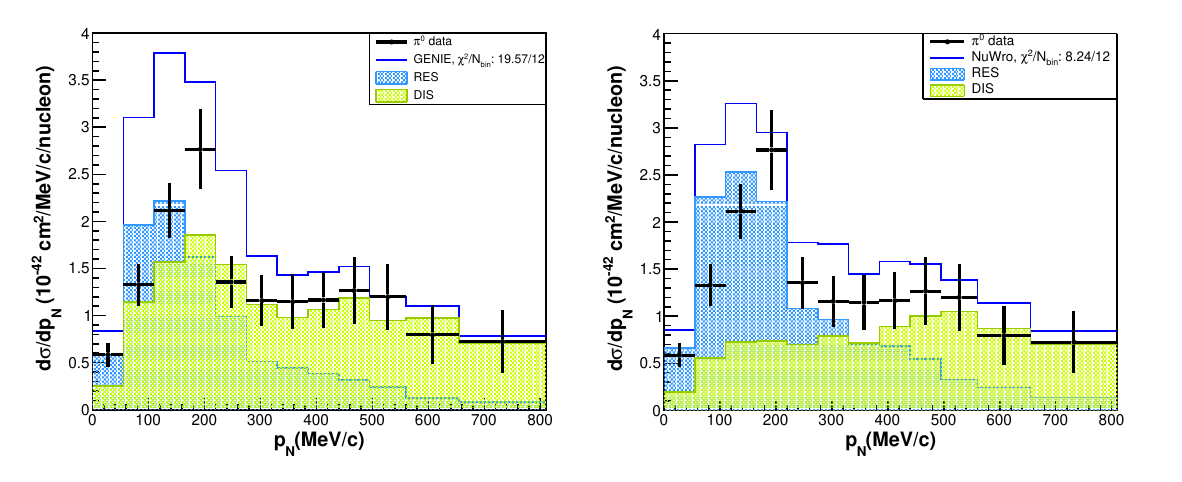}
    \caption{The flux integrated cross section in $p_N$ for single $\pi^0$ production using GENIE (left) and NuWro (Right) alternative models showing the contribution from RES and DIS.}
    \label{fig:4}
\end{figure*}

\subsection{Analysis using alternative MC models}

GENIE's predictions for LFG and SF are opposite to each other. However, LFG provides a better shape agreement than SF, so LFG is preferred for further GENIE analysis. The Fermi motion of the target nucleon is influenced by the nuclear ground state, as described by nuclear models. Therefore, the SRC fraction is considered an effective parameter and is set to 0.15 \cite{GENIE:2024ufm}. Since the RES channel contributes most to pion production, the RES axial mass is treated as a significant parameter and taken as 1.088962 \cite{GENIE:2024ufm}. The cross section in $p_N$ using the GENIE with the hN FSI model and the contributions of RES and DIS in single $\pi^0$ production are shown in the left panel of Fig.\ref{fig:4}. The prediction shows an improvement in the shape, but the Fermi peak still does not match the data. The $\chi^2$ per bin is also reduced to 19.57/12.  The higher prediction in the lower-momentum region is due to the higher DIS contribution in the lower momentum below the Fermi peak.

The higher contribution of DIS to the pion production causes the overprediction in NuWro. The FZ effect effectively affects the pion production \cite{Golan:2012wx}. 
It determines the amount of reinteractions of the produced hadrons and controls the strength of FSI. However, the FZ effect remains uncertain in the lower-energy range, and it is ambiguous whether it represents a physical phenomenon or just a parameter for FSI tuning \cite{golanphd2014}. NuWro considers the FZ effect for both particles created in the primary interaction and those produced during FSI. The FZ effect applied to the particles produced in the primary interaction is one of the reasons for the higher multiplicity of pions in the analysis \cite{Golan:2012wx}. The effect is disabled for primary particles in the analysis and only applied to the particles produced in FSI. The predictions for single $\pi^0$ production using the NuWro with LFG are shown in the right panel of Fig.\ref{fig:4}. There is an overprediction below the Fermi peak, similar to what is observed from GENIE. The $\chi^2$ per bin is also significantly reduced to 8.24/12 indicating an improvement in the prediction with a better overall shape. Neglecting the FZ effect for primary particles reduces the DIS contribution to single $\pi^0$ production compared to the default parameters, as shown in the right panel of Fig.\ref{fig:4}.

\section{\label{sec5} Conclusion}
 
In this paper, we investigate the struck neutron momentum in the MINER$\nu$A-$\pi^0$ cross section measurement on carbon nucleus using GENIE and NuWro models and consider the selection of 1$\mu$1p$1\pi^{0}$X$n$. From our analysis results, it can be observed that the default Monte Carlo parameters show deviation from the data with the 1$\mu$1p$1\pi^{0}$X$n$ selection alone. The prediction for MINER$\nu$A-$\pi^{0}$ using the alternative models matches the overall shape of the data and sustains reasonably well in the higher neutron momentum for both GENIE and NuWro, with the substantial outcome from NuWro in comparison with the default input. However, the RES contribution below the Fermi peak leaves no scope for the DIS component, leading to an overestimation in the lower momentum region. These studies show that our analysis with the alternative models significantly improved the event generators and it serves well in describing the $\pi^0$ data. This could be useful for future MINER$\nu$A measurements with higher statistics as well as future cross section measurements involving the pion production in both the accelerator and atmospheric experiments.    

\section*{Acknowledgements}

R. K. P. acknowledges the DST-INSPIRE Grant (2022/IF220293) for financial support. A. G. credits the grant support of the Department of Science and Technology (Grant No. SR/MF/PS-01/2016-IITH/G).

\bibliography{references}

\begin{thebibliography}{40}%
\makeatletter
\providecommand \@ifxundefined [1]{%
 \@ifx{#1\undefined}
}%
\providecommand \@ifnum [1]{%
 \ifnum #1\expandafter \@firstoftwo
 \else \expandafter \@secondoftwo
 \fi
}%
\providecommand \@ifx [1]{%
 \ifx #1\expandafter \@firstoftwo
 \else \expandafter \@secondoftwo
 \fi
}%
\providecommand \natexlab [1]{#1}%
\providecommand \enquote  [1]{``#1''}%
\providecommand \bibnamefont  [1]{#1}%
\providecommand \bibfnamefont [1]{#1}%
\providecommand \citenamefont [1]{#1}%
\providecommand \href@noop [0]{\@secondoftwo}%
\providecommand \href [0]{\begingroup \@sanitize@url \@href}%
\providecommand \@href[1]{\@@startlink{#1}\@@href}%
\providecommand \@@href[1]{\endgroup#1\@@endlink}%
\providecommand \@sanitize@url [0]{\catcode `\\12\catcode `\$12\catcode
  `\&12\catcode `\#12\catcode `\^12\catcode `\_12\catcode `\%12\relax}%
\providecommand \@@startlink[1]{}%
\providecommand \@@endlink[0]{}%
\providecommand \url  [0]{\begingroup\@sanitize@url \@url }%
\providecommand \@url [1]{\endgroup\@href {#1}{\urlprefix }}%
\providecommand \urlprefix  [0]{URL }%
\providecommand \Eprint [0]{\href }%
\providecommand \doibase [0]{https://doi.org/}%
\providecommand \selectlanguage [0]{\@gobble}%
\providecommand \bibinfo  [0]{\@secondoftwo}%
\providecommand \bibfield  [0]{\@secondoftwo}%
\providecommand \translation [1]{[#1]}%
\providecommand \BibitemOpen [0]{}%
\providecommand \bibitemStop [0]{}%
\providecommand \bibitemNoStop [0]{.\EOS\space}%
\providecommand \EOS [0]{\spacefactor3000\relax}%
\providecommand \BibitemShut  [1]{\csname bibitem#1\endcsname}%
\let\auto@bib@innerbib\@empty
\bibitem [{\citenamefont {Alvarez-Ruso}\ \emph {et~al.}(2018)\citenamefont
  {Alvarez-Ruso} \emph {et~al.}}]{NuSTEC:2017hzk}%
  \BibitemOpen
  \bibfield  {author} {\bibinfo {author} {\bibfnamefont {L.}~\bibnamefont
  {Alvarez-Ruso}} \emph {et~al.} (\bibinfo {collaboration} {NuSTEC}),\
  }\bibfield  {title} {\bibinfo {title} {{NuSTEC White Paper: Status and
  challenges of neutrino\textendash{}nucleus scattering}},\ }\href
  {https://doi.org/10.1016/j.ppnp.2018.01.006} {\bibfield  {journal} {\bibinfo
  {journal} {Prog. Part. Nucl. Phys.}\ }\textbf {\bibinfo {volume} {100}},\
  \bibinfo {pages} {1} (\bibinfo {year} {2018})},\ \Eprint
  {https://arxiv.org/abs/1706.03621} {arXiv:1706.03621 [hep-ph]} \BibitemShut
  {NoStop}%
\bibitem [{\citenamefont {Acero}\ \emph {et~al.}(2019)\citenamefont {Acero}
  \emph {et~al.}}]{NOvA:2019cyt}%
  \BibitemOpen
  \bibfield  {author} {\bibinfo {author} {\bibfnamefont {M.~A.}\ \bibnamefont
  {Acero}} \emph {et~al.} (\bibinfo {collaboration} {NOvA}),\ }\bibfield
  {title} {\bibinfo {title} {{First Measurement of Neutrino Oscillation
  Parameters using Neutrinos and Antineutrinos by NOvA}},\ }\href
  {https://doi.org/10.1103/PhysRevLett.123.151803} {\bibfield  {journal}
  {\bibinfo  {journal} {Phys. Rev. Lett.}\ }\textbf {\bibinfo {volume} {123}},\
  \bibinfo {pages} {151803} (\bibinfo {year} {2019})},\ \Eprint
  {https://arxiv.org/abs/1906.04907} {arXiv:1906.04907 [hep-ex]} \BibitemShut
  {NoStop}%
\bibitem [{\citenamefont {Abe}\ \emph {et~al.}(2020)\citenamefont {Abe} \emph
  {et~al.}}]{T2K:2019bcf}%
  \BibitemOpen
  \bibfield  {author} {\bibinfo {author} {\bibfnamefont {K.}~\bibnamefont
  {Abe}} \emph {et~al.} (\bibinfo {collaboration} {T2K}),\ }\bibfield  {title}
  {\bibinfo {title} {{Constraint on the matter\textendash{}antimatter
  symmetry-violating phase in neutrino oscillations}},\ }\href
  {https://doi.org/10.1038/s41586-020-2177-0} {\bibfield  {journal} {\bibinfo
  {journal} {Nature}\ }\textbf {\bibinfo {volume} {580}},\ \bibinfo {pages}
  {339} (\bibinfo {year} {2020})},\ \bibinfo {note} {[Erratum: Nature 583, E16
  (2020)]},\ \Eprint {https://arxiv.org/abs/1910.03887} {arXiv:1910.03887
  [hep-ex]} \BibitemShut {NoStop}%
\bibitem [{\citenamefont {Acciarri}\ \emph {et~al.}(2015)\citenamefont
  {Acciarri} \emph {et~al.}}]{DUNE:2015lol}%
  \BibitemOpen
  \bibfield  {author} {\bibinfo {author} {\bibfnamefont {R.}~\bibnamefont
  {Acciarri}} \emph {et~al.} (\bibinfo {collaboration} {DUNE}),\ }\bibfield
  {title} {\bibinfo {title} {{Long-Baseline Neutrino Facility (LBNF) and Deep
  Underground Neutrino Experiment (DUNE)}: {Conceptual Design Report, Volume 2:
  The Physics Program for DUNE at LBNF}},\ }\href@noop {} {\  (\bibinfo {year}
  {2015})},\ \Eprint {https://arxiv.org/abs/1512.06148} {arXiv:1512.06148
  [physics.ins-det]} \BibitemShut {NoStop}%
\bibitem [{\citenamefont {Lalnuntluanga}\ and\ \citenamefont
  {Giri}(2023)}]{Lalnuntluanga:2023qkp}%
  \BibitemOpen
  \bibfield  {author} {\bibinfo {author} {\bibfnamefont {R.}~\bibnamefont
  {Lalnuntluanga}}\ and\ \bibinfo {author} {\bibfnamefont {A.}~\bibnamefont
  {Giri}},\ }\bibfield  {title} {\bibinfo {title} {{Quantifying the second
  resonance effect in neutrino-Argon interaction using DUNE Near Detector}},\
  }\href {https://doi.org/10.1016/j.physletb.2023.137717} {\bibfield  {journal}
  {\bibinfo  {journal} {Phys. Lett. B}\ }\textbf {\bibinfo {volume} {838}},\
  \bibinfo {pages} {137717} (\bibinfo {year} {2023})}\BibitemShut {NoStop}%
\bibitem [{\citenamefont {Lalnuntluanga}\ and\ \citenamefont
  {Giri}(2024)}]{Lalnuntluanga:2024ssd}%
  \BibitemOpen
  \bibfield  {author} {\bibinfo {author} {\bibfnamefont {R.}~\bibnamefont
  {Lalnuntluanga}}\ and\ \bibinfo {author} {\bibfnamefont {A.}~\bibnamefont
  {Giri}},\ }\bibfield  {title} {\bibinfo {title} {{Pion Production in~DUNE
  Near Detector with~Argon Target}},\ }\href
  {https://doi.org/10.1007/978-981-97-0289-3_314} {\bibfield  {journal}
  {\bibinfo  {journal} {Springer Proc. Phys.}\ }\textbf {\bibinfo {volume}
  {304}},\ \bibinfo {pages} {1135} (\bibinfo {year} {2024})}\BibitemShut
  {NoStop}%
\bibitem [{\citenamefont {An}\ \emph {et~al.}(2016)\citenamefont {An} \emph
  {et~al.}}]{JUNO:2015zny}%
  \BibitemOpen
  \bibfield  {author} {\bibinfo {author} {\bibfnamefont {F.}~\bibnamefont {An}}
  \emph {et~al.} (\bibinfo {collaboration} {JUNO}),\ }\bibfield  {title}
  {\bibinfo {title} {{Neutrino Physics with JUNO}},\ }\href
  {https://doi.org/10.1088/0954-3899/43/3/030401} {\bibfield  {journal}
  {\bibinfo  {journal} {J. Phys. G}\ }\textbf {\bibinfo {volume} {43}},\
  \bibinfo {pages} {030401} (\bibinfo {year} {2016})},\ \Eprint
  {https://arxiv.org/abs/1507.05613} {arXiv:1507.05613 [physics.ins-det]}
  \BibitemShut {NoStop}%
\bibitem [{\citenamefont {Abe}\ \emph {et~al.}(2015)\citenamefont {Abe} \emph
  {et~al.}}]{Hyper-KamiokandeProto-:2015xww}%
  \BibitemOpen
  \bibfield  {author} {\bibinfo {author} {\bibfnamefont {K.}~\bibnamefont
  {Abe}} \emph {et~al.} (\bibinfo {collaboration} {Hyper-Kamiokande Proto-}),\
  }\bibfield  {title} {\bibinfo {title} {{Physics potential of a long-baseline
  neutrino oscillation experiment using a J-PARC neutrino beam and
  Hyper-Kamiokande}},\ }\href {https://doi.org/10.1093/ptep/ptv061} {\bibfield
  {journal} {\bibinfo  {journal} {PTEP}\ }\textbf {\bibinfo {volume} {2015}},\
  \bibinfo {pages} {053C02} (\bibinfo {year} {2015})},\ \Eprint
  {https://arxiv.org/abs/1502.05199} {arXiv:1502.05199 [hep-ex]} \BibitemShut
  {NoStop}%
\bibitem [{\citenamefont {Fukuda}\ \emph {et~al.}(1998)\citenamefont {Fukuda}
  \emph {et~al.}}]{Super-Kamiokande:1998kpq}%
  \BibitemOpen
  \bibfield  {author} {\bibinfo {author} {\bibfnamefont {Y.}~\bibnamefont
  {Fukuda}} \emph {et~al.} (\bibinfo {collaboration} {Super-Kamiokande}),\
  }\bibfield  {title} {\bibinfo {title} {{Evidence for oscillation of
  atmospheric neutrinos}},\ }\href
  {https://doi.org/10.1103/PhysRevLett.81.1562} {\bibfield  {journal} {\bibinfo
   {journal} {Phys. Rev. Lett.}\ }\textbf {\bibinfo {volume} {81}},\ \bibinfo
  {pages} {1562} (\bibinfo {year} {1998})},\ \Eprint
  {https://arxiv.org/abs/hep-ex/9807003} {arXiv:hep-ex/9807003} \BibitemShut
  {NoStop}%
\bibitem [{\citenamefont {Andreopoulos}\ \emph {et~al.}(2015)\citenamefont
  {Andreopoulos}, \citenamefont {Barry}, \citenamefont {Dytman}, \citenamefont
  {Gallagher}, \citenamefont {Golan}, \citenamefont {Hatcher}, \citenamefont
  {Perdue},\ and\ \citenamefont
  {Yarba}}]{andreopoulos2015genieneutrinomontecarlo}%
  \BibitemOpen
  \bibfield  {author} {\bibinfo {author} {\bibfnamefont {C.}~\bibnamefont
  {Andreopoulos}}, \bibinfo {author} {\bibfnamefont {C.}~\bibnamefont {Barry}},
  \bibinfo {author} {\bibfnamefont {S.}~\bibnamefont {Dytman}}, \bibinfo
  {author} {\bibfnamefont {H.}~\bibnamefont {Gallagher}}, \bibinfo {author}
  {\bibfnamefont {T.}~\bibnamefont {Golan}}, \bibinfo {author} {\bibfnamefont
  {R.}~\bibnamefont {Hatcher}}, \bibinfo {author} {\bibfnamefont
  {G.}~\bibnamefont {Perdue}},\ and\ \bibinfo {author} {\bibfnamefont
  {J.}~\bibnamefont {Yarba}},\ }\href {https://arxiv.org/abs/1510.05494}
  {\bibinfo {title} {The genie neutrino monte carlo generator: Physics and user
  manual}} (\bibinfo {year} {2015}),\ \Eprint
  {https://arxiv.org/abs/1510.05494} {arXiv:1510.05494 [hep-ph]} \BibitemShut
  {NoStop}%
\bibitem [{\citenamefont {Benhar}\ \emph {et~al.}(2005)\citenamefont {Benhar},
  \citenamefont {Farina}, \citenamefont {Nakamura}, \citenamefont {Sakuda},\
  and\ \citenamefont {Seki}}]{Benhar:2005dj}%
  \BibitemOpen
  \bibfield  {author} {\bibinfo {author} {\bibfnamefont {O.}~\bibnamefont
  {Benhar}}, \bibinfo {author} {\bibfnamefont {N.}~\bibnamefont {Farina}},
  \bibinfo {author} {\bibfnamefont {H.}~\bibnamefont {Nakamura}}, \bibinfo
  {author} {\bibfnamefont {M.}~\bibnamefont {Sakuda}},\ and\ \bibinfo {author}
  {\bibfnamefont {R.}~\bibnamefont {Seki}},\ }\bibfield  {title} {\bibinfo
  {title} {{Electron- and neutrino-nucleus scattering in the impulse
  approximation regime}},\ }\href {https://doi.org/10.1103/PhysRevD.72.053005}
  {\bibfield  {journal} {\bibinfo  {journal} {Phys. Rev. D}\ }\textbf {\bibinfo
  {volume} {72}},\ \bibinfo {pages} {053005} (\bibinfo {year} {2005})},\
  \Eprint {https://arxiv.org/abs/hep-ph/0506116} {arXiv:hep-ph/0506116}
  \BibitemShut {NoStop}%
\bibitem [{\citenamefont {Lu}\ \emph {et~al.}(2016)\citenamefont {Lu},
  \citenamefont {Pickering}, \citenamefont {Dolan}, \citenamefont {Barr},
  \citenamefont {Coplowe}, \citenamefont {Uchida}, \citenamefont {Wark},
  \citenamefont {Wascko}, \citenamefont {Weber},\ and\ \citenamefont
  {Yuan}}]{Lu:2015tcr}%
  \BibitemOpen
  \bibfield  {author} {\bibinfo {author} {\bibfnamefont {X.~G.}\ \bibnamefont
  {Lu}}, \bibinfo {author} {\bibfnamefont {L.}~\bibnamefont {Pickering}},
  \bibinfo {author} {\bibfnamefont {S.}~\bibnamefont {Dolan}}, \bibinfo
  {author} {\bibfnamefont {G.}~\bibnamefont {Barr}}, \bibinfo {author}
  {\bibfnamefont {D.}~\bibnamefont {Coplowe}}, \bibinfo {author} {\bibfnamefont
  {Y.}~\bibnamefont {Uchida}}, \bibinfo {author} {\bibfnamefont
  {D.}~\bibnamefont {Wark}}, \bibinfo {author} {\bibfnamefont {M.~O.}\
  \bibnamefont {Wascko}}, \bibinfo {author} {\bibfnamefont {A.}~\bibnamefont
  {Weber}},\ and\ \bibinfo {author} {\bibfnamefont {T.}~\bibnamefont {Yuan}},\
  }\bibfield  {title} {\bibinfo {title} {{Measurement of nuclear effects in
  neutrino interactions with minimal dependence on neutrino energy}},\ }\href
  {https://doi.org/10.1103/PhysRevC.94.015503} {\bibfield  {journal} {\bibinfo
  {journal} {Phys. Rev. C}\ }\textbf {\bibinfo {volume} {94}},\ \bibinfo
  {pages} {015503} (\bibinfo {year} {2016})},\ \Eprint
  {https://arxiv.org/abs/1512.05748} {arXiv:1512.05748 [nucl-th]} \BibitemShut
  {NoStop}%
\bibitem [{\citenamefont {Lu}\ \emph {et~al.}(2018)\citenamefont {Lu} \emph
  {et~al.}}]{MINERvA:2018hba}%
  \BibitemOpen
  \bibfield  {author} {\bibinfo {author} {\bibfnamefont {X.~G.}\ \bibnamefont
  {Lu}} \emph {et~al.} (\bibinfo {collaboration} {MINERvA}),\ }\bibfield
  {title} {\bibinfo {title} {{Measurement of final-state correlations in
  neutrino muon-proton mesonless production on hydrocarbon at $\langle
  E_\nu\rangle=3$ GeV}},\ }\href
  {https://doi.org/10.1103/PhysRevLett.121.022504} {\bibfield  {journal}
  {\bibinfo  {journal} {Phys. Rev. Lett.}\ }\textbf {\bibinfo {volume} {121}},\
  \bibinfo {pages} {022504} (\bibinfo {year} {2018})},\ \Eprint
  {https://arxiv.org/abs/1805.05486} {arXiv:1805.05486 [hep-ex]} \BibitemShut
  {NoStop}%
\bibitem [{\citenamefont {Cai}\ \emph {et~al.}(2020)\citenamefont {Cai} \emph
  {et~al.}}]{MINERvA:2019ope}%
  \BibitemOpen
  \bibfield  {author} {\bibinfo {author} {\bibfnamefont {T.}~\bibnamefont
  {Cai}} \emph {et~al.} (\bibinfo {collaboration} {MINERvA}),\ }\bibfield
  {title} {\bibinfo {title} {{Nucleon binding energy and transverse momentum
  imbalance in neutrino-nucleus reactions}},\ }\href
  {https://doi.org/10.1103/PhysRevD.101.092001} {\bibfield  {journal} {\bibinfo
   {journal} {Phys. Rev. D}\ }\textbf {\bibinfo {volume} {101}},\ \bibinfo
  {pages} {092001} (\bibinfo {year} {2020})},\ \Eprint
  {https://arxiv.org/abs/1910.08658} {arXiv:1910.08658 [hep-ex]} \BibitemShut
  {NoStop}%
\bibitem [{\citenamefont {Furmanski}\ and\ \citenamefont
  {Sobczyk}(2017)}]{Furmanski:2016wqo}%
  \BibitemOpen
  \bibfield  {author} {\bibinfo {author} {\bibfnamefont {A.~P.}\ \bibnamefont
  {Furmanski}}\ and\ \bibinfo {author} {\bibfnamefont {J.~T.}\ \bibnamefont
  {Sobczyk}},\ }\bibfield  {title} {\bibinfo {title} {{Neutrino energy
  reconstruction from one muon and one proton events}},\ }\href
  {https://doi.org/10.1103/PhysRevC.95.065501} {\bibfield  {journal} {\bibinfo
  {journal} {Phys. Rev. C}\ }\textbf {\bibinfo {volume} {95}},\ \bibinfo
  {pages} {065501} (\bibinfo {year} {2017})},\ \Eprint
  {https://arxiv.org/abs/1609.03530} {arXiv:1609.03530 [hep-ex]} \BibitemShut
  {NoStop}%
\bibitem [{\citenamefont {Abe}\ \emph {et~al.}(2018)\citenamefont {Abe} \emph
  {et~al.}}]{T2K:2018rnz}%
  \BibitemOpen
  \bibfield  {author} {\bibinfo {author} {\bibfnamefont {K.}~\bibnamefont
  {Abe}} \emph {et~al.} (\bibinfo {collaboration} {T2K}),\ }\bibfield  {title}
  {\bibinfo {title} {{Characterization of nuclear effects in muon-neutrino
  scattering on hydrocarbon with a measurement of final-state kinematics and
  correlations in charged-current pionless interactions at T2K}},\ }\href
  {https://doi.org/10.1103/PhysRevD.98.032003} {\bibfield  {journal} {\bibinfo
  {journal} {Phys. Rev. D}\ }\textbf {\bibinfo {volume} {98}},\ \bibinfo
  {pages} {032003} (\bibinfo {year} {2018})},\ \Eprint
  {https://arxiv.org/abs/1802.05078} {arXiv:1802.05078 [hep-ex]} \BibitemShut
  {NoStop}%
\bibitem [{\citenamefont {Cai}\ \emph {et~al.}(2019)\citenamefont {Cai},
  \citenamefont {Lu},\ and\ \citenamefont {Ruterbories}}]{Cai:2019jzk}%
  \BibitemOpen
  \bibfield  {author} {\bibinfo {author} {\bibfnamefont {T.}~\bibnamefont
  {Cai}}, \bibinfo {author} {\bibfnamefont {X.}~\bibnamefont {Lu}},\ and\
  \bibinfo {author} {\bibfnamefont {D.}~\bibnamefont {Ruterbories}},\
  }\bibfield  {title} {\bibinfo {title} {{Pion-proton correlation in neutrino
  interactions on nuclei}},\ }\href
  {https://doi.org/10.1103/PhysRevD.100.073010} {\bibfield  {journal} {\bibinfo
   {journal} {Phys. Rev. D}\ }\textbf {\bibinfo {volume} {100}},\ \bibinfo
  {pages} {073010} (\bibinfo {year} {2019})},\ \Eprint
  {https://arxiv.org/abs/1907.11212} {arXiv:1907.11212 [hep-ex]} \BibitemShut
  {NoStop}%
\bibitem [{\citenamefont {Coplowe}\ \emph {et~al.}(2020)\citenamefont {Coplowe}
  \emph {et~al.}}]{MINERvA:2020anu}%
  \BibitemOpen
  \bibfield  {author} {\bibinfo {author} {\bibfnamefont {D.}~\bibnamefont
  {Coplowe}} \emph {et~al.} (\bibinfo {collaboration} {MINERvA}),\ }\bibfield
  {title} {\bibinfo {title} {{Probing nuclear effects with neutrino-induced
  charged-current neutral pion production}},\ }\href
  {https://doi.org/10.1103/PhysRevD.102.072007} {\bibfield  {journal} {\bibinfo
   {journal} {Phys. Rev. D}\ }\textbf {\bibinfo {volume} {102}},\ \bibinfo
  {pages} {072007} (\bibinfo {year} {2020})},\ \Eprint
  {https://arxiv.org/abs/2002.05812} {arXiv:2002.05812 [hep-ex]} \BibitemShut
  {NoStop}%
\bibitem [{\citenamefont {Buss}\ \emph {et~al.}(2012)\citenamefont {Buss},
  \citenamefont {Gaitanos}, \citenamefont {Gallmeister}, \citenamefont {van
  Hees}, \citenamefont {Kaskulov}, \citenamefont {Lalakulich}, \citenamefont
  {Larionov}, \citenamefont {Leitner}, \citenamefont {Weil},\ and\
  \citenamefont {Mosel}}]{Buss:2011mx}%
  \BibitemOpen
  \bibfield  {author} {\bibinfo {author} {\bibfnamefont {O.}~\bibnamefont
  {Buss}}, \bibinfo {author} {\bibfnamefont {T.}~\bibnamefont {Gaitanos}},
  \bibinfo {author} {\bibfnamefont {K.}~\bibnamefont {Gallmeister}}, \bibinfo
  {author} {\bibfnamefont {H.}~\bibnamefont {van Hees}}, \bibinfo {author}
  {\bibfnamefont {M.}~\bibnamefont {Kaskulov}}, \bibinfo {author}
  {\bibfnamefont {O.}~\bibnamefont {Lalakulich}}, \bibinfo {author}
  {\bibfnamefont {A.~B.}\ \bibnamefont {Larionov}}, \bibinfo {author}
  {\bibfnamefont {T.}~\bibnamefont {Leitner}}, \bibinfo {author} {\bibfnamefont
  {J.}~\bibnamefont {Weil}},\ and\ \bibinfo {author} {\bibfnamefont
  {U.}~\bibnamefont {Mosel}},\ }\bibfield  {title} {\bibinfo {title}
  {{Transport-theoretical Description of Nuclear Reactions}},\ }\href
  {https://doi.org/10.1016/j.physrep.2011.12.001} {\bibfield  {journal}
  {\bibinfo  {journal} {Phys. Rept.}\ }\textbf {\bibinfo {volume} {512}},\
  \bibinfo {pages} {1} (\bibinfo {year} {2012})},\ \Eprint
  {https://arxiv.org/abs/1106.1344} {arXiv:1106.1344 [hep-ph]} \BibitemShut
  {NoStop}%
\bibitem [{\citenamefont {Yan}\ \emph {et~al.}(2024)\citenamefont {Yan},
  \citenamefont {Niewczas}, \citenamefont {Nikolakopoulos}, \citenamefont
  {Gonz\'alez-Jim\'enez}, \citenamefont {Jachowicz}, \citenamefont {Lu},
  \citenamefont {Sobczyk},\ and\ \citenamefont {Zheng}}]{Yan:2024kkg}%
  \BibitemOpen
  \bibfield  {author} {\bibinfo {author} {\bibfnamefont {Q.}~\bibnamefont
  {Yan}}, \bibinfo {author} {\bibfnamefont {K.}~\bibnamefont {Niewczas}},
  \bibinfo {author} {\bibfnamefont {A.}~\bibnamefont {Nikolakopoulos}},
  \bibinfo {author} {\bibfnamefont {R.}~\bibnamefont {Gonz\'alez-Jim\'enez}},
  \bibinfo {author} {\bibfnamefont {N.}~\bibnamefont {Jachowicz}}, \bibinfo
  {author} {\bibfnamefont {X.}~\bibnamefont {Lu}}, \bibinfo {author}
  {\bibfnamefont {J.}~\bibnamefont {Sobczyk}},\ and\ \bibinfo {author}
  {\bibfnamefont {Y.}~\bibnamefont {Zheng}},\ }\bibfield  {title} {\bibinfo
  {title} {{The Ghent Hybrid Model in NuWro: a new neutrino single-pion
  production model in the GeV regime}},\ }\href@noop {} {\  (\bibinfo {year}
  {2024})},\ \Eprint {https://arxiv.org/abs/2405.05212} {arXiv:2405.05212
  [hep-ph]} \BibitemShut {NoStop}%
\bibitem [{\citenamefont {Li}\ \emph {et~al.}(2024)\citenamefont {Li} \emph
  {et~al.}}]{GENIE:2024ufm}%
  \BibitemOpen
  \bibfield  {author} {\bibinfo {author} {\bibfnamefont {W.}~\bibnamefont {Li}}
  \emph {et~al.} (\bibinfo {collaboration} {GENIE}),\ }\bibfield  {title}
  {\bibinfo {title} {{First combined tuning on transverse kinematic imbalance
  data with and without pion production constraints}},\ }\href@noop {} {\
  (\bibinfo {year} {2024})},\ \Eprint {https://arxiv.org/abs/2404.08510}
  {arXiv:2404.08510 [hep-ex]} \BibitemShut {NoStop}%
\bibitem [{\citenamefont {Juszczak}\ \emph {et~al.}(2006)\citenamefont
  {Juszczak}, \citenamefont {Nowak},\ and\ \citenamefont
  {Sobczyk}}]{Juszczak:2005zs}%
  \BibitemOpen
  \bibfield  {author} {\bibinfo {author} {\bibfnamefont {C.}~\bibnamefont
  {Juszczak}}, \bibinfo {author} {\bibfnamefont {J.~A.}\ \bibnamefont
  {Nowak}},\ and\ \bibinfo {author} {\bibfnamefont {J.~T.}\ \bibnamefont
  {Sobczyk}},\ }\bibfield  {title} {\bibinfo {title} {{Simulations from a new
  neutrino event generator}},\ }\href
  {https://doi.org/10.1016/j.nuclphysbps.2006.08.069} {\bibfield  {journal}
  {\bibinfo  {journal} {Nucl. Phys. B Proc. Suppl.}\ }\textbf {\bibinfo
  {volume} {159}},\ \bibinfo {pages} {211} (\bibinfo {year} {2006})},\ \Eprint
  {https://arxiv.org/abs/hep-ph/0512365} {arXiv:hep-ph/0512365} \BibitemShut
  {NoStop}%
\bibitem [{\citenamefont {Aliaga}\ \emph {et~al.}(2016)\citenamefont {Aliaga}
  \emph {et~al.}}]{MINERvA:2016iqn}%
  \BibitemOpen
  \bibfield  {author} {\bibinfo {author} {\bibfnamefont {L.}~\bibnamefont
  {Aliaga}} \emph {et~al.} (\bibinfo {collaboration} {MINERvA}),\ }\bibfield
  {title} {\bibinfo {title} {{Neutrino Flux Predictions for the NuMI Beam}},\
  }\href {https://doi.org/10.1103/PhysRevD.94.092005} {\bibfield  {journal}
  {\bibinfo  {journal} {Phys. Rev. D}\ }\textbf {\bibinfo {volume} {94}},\
  \bibinfo {pages} {092005} (\bibinfo {year} {2016})},\ \bibinfo {note}
  {[Addendum: Phys.Rev.D 95, 039903 (2017)]},\ \Eprint
  {https://arxiv.org/abs/1607.00704} {arXiv:1607.00704 [hep-ex]} \BibitemShut
  {NoStop}%
\bibitem [{\citenamefont {Benhar}\ \emph {et~al.}(1994)\citenamefont {Benhar},
  \citenamefont {Fabrocini}, \citenamefont {Fantoni},\ and\ \citenamefont
  {Sick}}]{Benhar:1994hw}%
  \BibitemOpen
  \bibfield  {author} {\bibinfo {author} {\bibfnamefont {O.}~\bibnamefont
  {Benhar}}, \bibinfo {author} {\bibfnamefont {A.}~\bibnamefont {Fabrocini}},
  \bibinfo {author} {\bibfnamefont {S.}~\bibnamefont {Fantoni}},\ and\ \bibinfo
  {author} {\bibfnamefont {I.}~\bibnamefont {Sick}},\ }\bibfield  {title}
  {\bibinfo {title} {{Spectral function of finite nuclei and scattering of GeV
  electrons}},\ }\href {https://doi.org/10.1016/0375-9474(94)90920-2}
  {\bibfield  {journal} {\bibinfo  {journal} {Nucl. Phys. A}\ }\textbf
  {\bibinfo {volume} {579}},\ \bibinfo {pages} {493} (\bibinfo {year}
  {1994})}\BibitemShut {NoStop}%
\bibitem [{\citenamefont {Cruz-Torres}\ \emph {et~al.}(2018)\citenamefont
  {Cruz-Torres}, \citenamefont {Schmidt}, \citenamefont {Miller}, \citenamefont
  {Weinstein}, \citenamefont {Barnea}, \citenamefont {Weiss}, \citenamefont
  {Piasetzky},\ and\ \citenamefont {Hen}}]{Cruz-Torres:2017sjy}%
  \BibitemOpen
  \bibfield  {author} {\bibinfo {author} {\bibfnamefont {R.}~\bibnamefont
  {Cruz-Torres}}, \bibinfo {author} {\bibfnamefont {A.}~\bibnamefont
  {Schmidt}}, \bibinfo {author} {\bibfnamefont {G.~A.}\ \bibnamefont {Miller}},
  \bibinfo {author} {\bibfnamefont {L.~B.}\ \bibnamefont {Weinstein}}, \bibinfo
  {author} {\bibfnamefont {N.}~\bibnamefont {Barnea}}, \bibinfo {author}
  {\bibfnamefont {R.}~\bibnamefont {Weiss}}, \bibinfo {author} {\bibfnamefont
  {E.}~\bibnamefont {Piasetzky}},\ and\ \bibinfo {author} {\bibfnamefont
  {O.}~\bibnamefont {Hen}},\ }\bibfield  {title} {\bibinfo {title} {{Short
  range correlations and the isospin dependence of nuclear correlation
  functions}},\ }\href {https://doi.org/10.1016/j.physletb.2018.07.069}
  {\bibfield  {journal} {\bibinfo  {journal} {Phys. Lett. B}\ }\textbf
  {\bibinfo {volume} {785}},\ \bibinfo {pages} {304} (\bibinfo {year}
  {2018})},\ \Eprint {https://arxiv.org/abs/1710.07966} {arXiv:1710.07966
  [nucl-th]} \BibitemShut {NoStop}%
\bibitem [{\citenamefont {Gran}\ \emph {et~al.}(2013)\citenamefont {Gran},
  \citenamefont {Nieves}, \citenamefont {Sanchez},\ and\ \citenamefont
  {Vicente~Vacas}}]{Gran:2013kda}%
  \BibitemOpen
  \bibfield  {author} {\bibinfo {author} {\bibfnamefont {R.}~\bibnamefont
  {Gran}}, \bibinfo {author} {\bibfnamefont {J.}~\bibnamefont {Nieves}},
  \bibinfo {author} {\bibfnamefont {F.}~\bibnamefont {Sanchez}},\ and\ \bibinfo
  {author} {\bibfnamefont {M.~J.}\ \bibnamefont {Vicente~Vacas}},\ }\bibfield
  {title} {\bibinfo {title} {{Neutrino-nucleus quasi-elastic and 2p2h
  interactions up to 10 GeV}},\ }\href
  {https://doi.org/10.1103/PhysRevD.88.113007} {\bibfield  {journal} {\bibinfo
  {journal} {Phys. Rev. D}\ }\textbf {\bibinfo {volume} {88}},\ \bibinfo
  {pages} {113007} (\bibinfo {year} {2013})},\ \Eprint
  {https://arxiv.org/abs/1307.8105} {arXiv:1307.8105 [hep-ph]} \BibitemShut
  {NoStop}%
\bibitem [{\citenamefont {Nieves}\ \emph {et~al.}(2016)\citenamefont {Nieves},
  \citenamefont {Simo}, \citenamefont {S\'anchez},\ and\ \citenamefont
  {Vicente~Vacas}}]{Nieves:2016sma}%
  \BibitemOpen
  \bibfield  {author} {\bibinfo {author} {\bibfnamefont {J.}~\bibnamefont
  {Nieves}}, \bibinfo {author} {\bibfnamefont {I.~R.}\ \bibnamefont {Simo}},
  \bibinfo {author} {\bibfnamefont {F.}~\bibnamefont {S\'anchez}},\ and\
  \bibinfo {author} {\bibfnamefont {M.~J.}\ \bibnamefont {Vicente~Vacas}},\
  }\bibfield  {title} {\bibinfo {title} {{2p2h Excitations, MEC, Nucleon
  Correlations and Other Sources of QE-like Events}},\ }\href
  {https://doi.org/10.7566/JPSCP.12.010002} {\bibfield  {journal} {\bibinfo
  {journal} {JPS Conf. Proc.}\ }\textbf {\bibinfo {volume} {12}},\ \bibinfo
  {pages} {010002} (\bibinfo {year} {2016})}\BibitemShut {NoStop}%
\bibitem [{\citenamefont {Berger}\ and\ \citenamefont
  {Sehgal}(2007)}]{Berger:2007rq}%
  \BibitemOpen
  \bibfield  {author} {\bibinfo {author} {\bibfnamefont {C.}~\bibnamefont
  {Berger}}\ and\ \bibinfo {author} {\bibfnamefont {L.~M.}\ \bibnamefont
  {Sehgal}},\ }\bibfield  {title} {\bibinfo {title} {{Lepton mass effects in
  single pion production by neutrinos}},\ }\href
  {https://doi.org/10.1103/PhysRevD.76.113004} {\bibfield  {journal} {\bibinfo
  {journal} {Phys. Rev. D}\ }\textbf {\bibinfo {volume} {76}},\ \bibinfo
  {pages} {113004} (\bibinfo {year} {2007})},\ \Eprint
  {https://arxiv.org/abs/0709.4378} {arXiv:0709.4378 [hep-ph]} \BibitemShut
  {NoStop}%
\bibitem [{\citenamefont {Bodek}\ and\ \citenamefont
  {Yang}(2002)}]{Bodek:2002vp}%
  \BibitemOpen
  \bibfield  {author} {\bibinfo {author} {\bibfnamefont {A.}~\bibnamefont
  {Bodek}}\ and\ \bibinfo {author} {\bibfnamefont {U.~K.}\ \bibnamefont
  {Yang}},\ }\bibfield  {title} {\bibinfo {title} {{Modeling deep inelastic
  cross-sections in the few GeV region}},\ }\href
  {https://doi.org/10.1016/S0920-5632(02)01755-3} {\bibfield  {journal}
  {\bibinfo  {journal} {Nucl. Phys. B Proc. Suppl.}\ }\textbf {\bibinfo
  {volume} {112}},\ \bibinfo {pages} {70} (\bibinfo {year} {2002})},\ \Eprint
  {https://arxiv.org/abs/hep-ex/0203009} {arXiv:hep-ex/0203009} \BibitemShut
  {NoStop}%
\bibitem [{\citenamefont {Alvarez-Ruso}\ \emph {et~al.}(2021)\citenamefont
  {Alvarez-Ruso} \emph {et~al.}}]{GENIE:2021npt}%
  \BibitemOpen
  \bibfield  {author} {\bibinfo {author} {\bibfnamefont {L.}~\bibnamefont
  {Alvarez-Ruso}} \emph {et~al.} (\bibinfo {collaboration} {GENIE}),\
  }\bibfield  {title} {\bibinfo {title} {{Recent highlights from GENIE v3}},\
  }\href {https://doi.org/10.1140/epjs/s11734-021-00295-7} {\bibfield
  {journal} {\bibinfo  {journal} {Eur. Phys. J. ST}\ }\textbf {\bibinfo
  {volume} {230}},\ \bibinfo {pages} {4449} (\bibinfo {year} {2021})},\ \Eprint
  {https://arxiv.org/abs/2106.09381} {arXiv:2106.09381 [hep-ph]} \BibitemShut
  {NoStop}%
\bibitem [{\citenamefont {Meng}\ \emph {et~al.}(2023)\citenamefont {Meng},
  \citenamefont {Lu},\ and\ \citenamefont {Xu}}]{Meng:2023age}%
  \BibitemOpen
  \bibfield  {author} {\bibinfo {author} {\bibfnamefont {Q.}~\bibnamefont
  {Meng}}, \bibinfo {author} {\bibfnamefont {Z.}~\bibnamefont {Lu}},\ and\
  \bibinfo {author} {\bibfnamefont {C.}~\bibnamefont {Xu}},\ }\bibfield
  {title} {\bibinfo {title} {{Short-range correlations and momentum
  distributions in mirror nuclei H3 and He3}},\ }\href
  {https://doi.org/10.1103/PhysRevC.108.014001} {\bibfield  {journal} {\bibinfo
   {journal} {Phys. Rev. C}\ }\textbf {\bibinfo {volume} {108}},\ \bibinfo
  {pages} {014001} (\bibinfo {year} {2023})},\ \Eprint
  {https://arxiv.org/abs/2307.14592} {arXiv:2307.14592 [nucl-th]} \BibitemShut
  {NoStop}%
\bibitem [{\citenamefont {Guo}\ \emph {et~al.}(2021)\citenamefont {Guo},
  \citenamefont {Li},\ and\ \citenamefont {Yong}}]{Guo:2021zcs}%
  \BibitemOpen
  \bibfield  {author} {\bibinfo {author} {\bibfnamefont {W.-M.}\ \bibnamefont
  {Guo}}, \bibinfo {author} {\bibfnamefont {B.-A.}\ \bibnamefont {Li}},\ and\
  \bibinfo {author} {\bibfnamefont {G.-C.}\ \bibnamefont {Yong}},\ }\bibfield
  {title} {\bibinfo {title} {{Imprints of high-momentum nucleons in nuclei on
  hard photons from heavy-ion collisions near the Fermi energy}},\ }\href
  {https://doi.org/10.1103/PhysRevC.104.034603} {\bibfield  {journal} {\bibinfo
   {journal} {Phys. Rev. C}\ }\textbf {\bibinfo {volume} {104}},\ \bibinfo
  {pages} {034603} (\bibinfo {year} {2021})},\ \Eprint
  {https://arxiv.org/abs/2106.08242} {arXiv:2106.08242 [nucl-th]} \BibitemShut
  {NoStop}%
\bibitem [{\citenamefont {Egiyan}\ \emph {et~al.}(2006)\citenamefont {Egiyan}
  \emph {et~al.}}]{CLAS:2005ola}%
  \BibitemOpen
  \bibfield  {author} {\bibinfo {author} {\bibfnamefont {K.~S.}\ \bibnamefont
  {Egiyan}} \emph {et~al.} (\bibinfo {collaboration} {CLAS}),\ }\bibfield
  {title} {\bibinfo {title} {{Measurement of 2- and 3-nucleon short range
  correlation probabilities in nuclei}},\ }\href
  {https://doi.org/10.1103/PhysRevLett.96.082501} {\bibfield  {journal}
  {\bibinfo  {journal} {Phys. Rev. Lett.}\ }\textbf {\bibinfo {volume} {96}},\
  \bibinfo {pages} {082501} (\bibinfo {year} {2006})},\ \Eprint
  {https://arxiv.org/abs/nucl-ex/0508026} {arXiv:nucl-ex/0508026} \BibitemShut
  {NoStop}%
\bibitem [{\citenamefont {Llewellyn~Smith}(1972)}]{LlewellynSmith:1971uhs}%
  \BibitemOpen
  \bibfield  {author} {\bibinfo {author} {\bibfnamefont {C.~H.}\ \bibnamefont
  {Llewellyn~Smith}},\ }\bibfield  {title} {\bibinfo {title} {{Neutrino
  Reactions at Accelerator Energies}},\ }\href
  {https://doi.org/10.1016/0370-1573(72)90010-5} {\bibfield  {journal}
  {\bibinfo  {journal} {Phys. Rept.}\ }\textbf {\bibinfo {volume} {3}},\
  \bibinfo {pages} {261} (\bibinfo {year} {1972})}\BibitemShut {NoStop}%
\bibitem [{\citenamefont {Rein}\ and\ \citenamefont
  {Sehgal}(1981)}]{Rein:1980wg}%
  \BibitemOpen
  \bibfield  {author} {\bibinfo {author} {\bibfnamefont {D.}~\bibnamefont
  {Rein}}\ and\ \bibinfo {author} {\bibfnamefont {L.~M.}\ \bibnamefont
  {Sehgal}},\ }\bibfield  {title} {\bibinfo {title} {{Neutrino Excitation of
  Baryon Resonances and Single Pion Production}},\ }\href
  {https://doi.org/10.1016/0003-4916(81)90242-6} {\bibfield  {journal}
  {\bibinfo  {journal} {Annals Phys.}\ }\textbf {\bibinfo {volume} {133}},\
  \bibinfo {pages} {79} (\bibinfo {year} {1981})}\BibitemShut {NoStop}%
\bibitem [{\citenamefont {Metropolis}\ \emph {et~al.}(1958)\citenamefont
  {Metropolis}, \citenamefont {Bivins}, \citenamefont {Storm}, \citenamefont
  {Miller}, \citenamefont {Friedlander},\ and\ \citenamefont
  {Turkevich}}]{Metropolis:1958sb}%
  \BibitemOpen
  \bibfield  {author} {\bibinfo {author} {\bibfnamefont {N.}~\bibnamefont
  {Metropolis}}, \bibinfo {author} {\bibfnamefont {R.}~\bibnamefont {Bivins}},
  \bibinfo {author} {\bibfnamefont {M.}~\bibnamefont {Storm}}, \bibinfo
  {author} {\bibfnamefont {J.~M.}\ \bibnamefont {Miller}}, \bibinfo {author}
  {\bibfnamefont {G.}~\bibnamefont {Friedlander}},\ and\ \bibinfo {author}
  {\bibfnamefont {A.}~\bibnamefont {Turkevich}},\ }\bibfield  {title} {\bibinfo
  {title} {{Monte Carlo Calculations on Intranuclear Cascades. 2. High-Energy
  Studies and Pion Processes}},\ }\href
  {https://doi.org/10.1103/PhysRev.110.204} {\bibfield  {journal} {\bibinfo
  {journal} {Phys. Rev.}\ }\textbf {\bibinfo {volume} {110}},\ \bibinfo {pages}
  {204} (\bibinfo {year} {1958})}\BibitemShut {NoStop}%
\bibitem [{\citenamefont {Golan}\ \emph {et~al.}(2012)\citenamefont {Golan},
  \citenamefont {Juszczak},\ and\ \citenamefont {Sobczyk}}]{Golan:2012wx}%
  \BibitemOpen
  \bibfield  {author} {\bibinfo {author} {\bibfnamefont {T.}~\bibnamefont
  {Golan}}, \bibinfo {author} {\bibfnamefont {C.}~\bibnamefont {Juszczak}},\
  and\ \bibinfo {author} {\bibfnamefont {J.~T.}\ \bibnamefont {Sobczyk}},\
  }\bibfield  {title} {\bibinfo {title} {{Final State Interactions Effects in
  Neutrino-Nucleus Interactions}},\ }\href
  {https://doi.org/10.1103/PhysRevC.86.015505} {\bibfield  {journal} {\bibinfo
  {journal} {Phys. Rev. C}\ }\textbf {\bibinfo {volume} {86}},\ \bibinfo
  {pages} {015505} (\bibinfo {year} {2012})},\ \Eprint
  {https://arxiv.org/abs/1202.4197} {arXiv:1202.4197 [nucl-th]} \BibitemShut
  {NoStop}%
\bibitem [{\citenamefont {Battistoni}\ \emph {et~al.}(2009)\citenamefont
  {Battistoni}, \citenamefont {Sala}, \citenamefont {Lantz}, \citenamefont
  {Ferrari},\ and\ \citenamefont {Smirnov}}]{Battistoni:2009zzb}%
  \BibitemOpen
  \bibfield  {author} {\bibinfo {author} {\bibfnamefont {G.}~\bibnamefont
  {Battistoni}}, \bibinfo {author} {\bibfnamefont {P.~R.}\ \bibnamefont
  {Sala}}, \bibinfo {author} {\bibfnamefont {M.}~\bibnamefont {Lantz}},
  \bibinfo {author} {\bibfnamefont {A.}~\bibnamefont {Ferrari}},\ and\ \bibinfo
  {author} {\bibfnamefont {G.}~\bibnamefont {Smirnov}},\ }\bibfield  {title}
  {\bibinfo {title} {{Neutrino interactions with FLUKA}},\ }\href
  {https://www.actaphys.uj.edu.pl/R/40/9/2491/pdf} {\bibfield  {journal}
  {\bibinfo  {journal} {Acta Phys. Polon. B}\ }\textbf {\bibinfo {volume}
  {40}},\ \bibinfo {pages} {2491} (\bibinfo {year} {2009})}\BibitemShut
  {NoStop}%
\bibitem [{\citenamefont {Ranft}(1989)}]{Ranft:1988kc}%
  \BibitemOpen
  \bibfield  {author} {\bibinfo {author} {\bibfnamefont {J.}~\bibnamefont
  {Ranft}},\ }\bibfield  {title} {\bibinfo {title} {{Hadron Production in
  Hadron - Nucleus and Nucleus-nucleus Collisions in a Dual Parton Model
  Modified by a Formation Zone Intranuclear Cascade}},\ }\href
  {https://doi.org/10.1007/BF01506540} {\bibfield  {journal} {\bibinfo
  {journal} {Z. Phys. C}\ }\textbf {\bibinfo {volume} {43}},\ \bibinfo {pages}
  {439} (\bibinfo {year} {1989})}\BibitemShut {NoStop}%
\bibitem [{\citenamefont {Golan}(2014)}]{golanphd2014}%
  \BibitemOpen
  \bibfield  {author} {\bibinfo {author} {\bibfnamefont {T.}~\bibnamefont
  {Golan}},\ }\emph {\bibinfo {title} {Modeling nuclear effects in NuWro Monte
  Carlo neutrino event generator}},\ \href
  {http://wng.ift.uni.wroc.pl/files/Golan_PhD.pdf} {Ph.D. thesis},\ \bibinfo
  {school} {University of Wrocław, Poland} (\bibinfo {year}
  {2014})\BibitemShut {NoStop}%
\end{thebibliography}%

\end{document}